\documentstyle[12pt]{report}

\def \kC {K^+_o({\cal C})}
\def \pC {Pic({\cal C})}
\def \gC {Gr({\cal C})}
\def \nC {\otimes\! -\! Nat(id_{\cal C})}
\def \rnC {Nat(id_{\cal C})}
\def  \bZ {{\bf Z}}
\def\twoheadrightarrow{\to\mkern-20mu\to}
\def\thrafill{$\mathsurround=0pt \mathord- \mkern-6mu
\cleaders\hbox{$\mkern-2mu
\mathord- \mkern-2mu$}\hfill \mkern-6mu\mathord\twoheadrightarrow$}
\mathchardef\ll="3015
\def \hup {\hbox{$\Biggl{\uparrow}$}
\mkern-11.5mu\raise-20.2pt\hbox{$\ll$}}
\def\bC {{\bf C}}
\def \bR {{\bf R}}

\newtheorem{theorem}{Theorem} [chapter]

\begin{document}
\begin{titlepage}

\begin{center}

\section*{On Braided Tensorcategories}

\vspace*{1.5cm}
\bigskip

{\large Thomas Kerler\\
 \medskip
Department of Mathematics\\
Harvard University\\
Cambridge, MA, USA}\\
kerler@math.harvard.edu\\

\end{center}
\vspace*{3cm}

{\small Abstract\footnote{Submitted as contribution to
the Proccedings of the XXIIth International Conference on Differential
Geometric Methods in Theoretical Physics, Ixtapa-Zihuatanejo, M\'exico. Kluwer
Academic Publishers, 1993.}
: We investigate invertible elements and gradings in braided tensor categories.
This leads us to the definition of theta-, product-, subgrading and
orbitcategories in order to construct new families of BTC's from given ones.
We use the representation theory of Hecke algebras in order to relate the
fusionring of a BTC generated by an object  $X$ with a two component
decomposition of
its tensorsquare to the fusionring of quantum groups of type $A$ at roots of
unity. We find the condition of `local isomorphie' on a special fusionring
morphism implying that a BTC is
obtained from the above constructions applied to the semisimplified
representation category of a quantum group. This family of BTC's contains new
series of twisted categories that do not stem from known Hopf algebras.
Using the language of incidence graphs and
the balancing structure on a BTC we also find strong constraints on the
fusionring morphism. For Temperley Lieb type categories these are sufficient
to show local isomorphie. Thus we obtain a  classification for the subclass of
Temperley Lieb type categories.}

\end{titlepage}

\newpage

%\section*{On Braided Tensorcategories}

%\noindent
%Thomas Kerler\\
%Department of Mathematics\\
%Harvard University\\
%Cambridge, MA, USA\\
%kerler@math.harvard.edu\\

\setcounter{chapter}{0}
\subsection*{0.Introduction}
 An important step in
organizing selection rules and defining symmetry principles of Quantumtheories
in algebraic terms has been the introduction of group theory into physics by
Weyl,Yang, Mills and others. Since the works
of [DR] and [D] it has become clear that the relevant data can be equivalently
and more directly described by a {\em symmetric tensorcategory} (STC). Often in
low dimensional physics the axiom that the commutativity constraint squares to
one has to be relaxed so that we naturally obtain representation of the braid
groups rather than the symmetric groups.
The more general {\em braided tensorcategories} (BTC) are related to
quasitriangular quasi-Hopfalgebras, but there is no one to one
duality-correspondence as for STC's since BTC's are rarely Tannakian.
Interestingly, they appear in many other areas of mathematical physics like the
theory of subfactors of von Neumann algebras, two dimensional integrable
lattice models, and low dimensional topology.

At generic points in the space of BTC's many uniqueness statements can be found
by using deformation theory. They give some explanation about the relation of
affine algebras and quantum groups at generic levels. For {\em rational}
theories these methods break down.
Nevertheless, one has identified equivalent rational BTC's coming from very
different areas. An example of a family of related rational models
includes $SU(2)$ and rank=2 WZW-models, the corresponding quantum groups at
roots of unity, subfactors with Jones-index $< 4\,$, the  Alexander or Jones
polynomials, and the $Q$-state Pottsmodel. In order to explain these
coincidences in terms of a classification we need to find reasonable
constraints on the considered class of BTC's. Most conveniently they are
imposed on the combinatorial part of the $\otimes$-category, i.e., the
fusionrules.

In [KW] (see also [FK] for $k=2$) it has been shown that if the entire
fusionring
of a BTC  is equal to that of $\overline{Rep}\bigl(U_q(sl(k))\bigr)$ then the
two categories themselves have to be isomorphic for suitable $q$. In this paper
(which is in large parts a summary of results from [FK]) we wish to impose a
much weaker condition, namely that the category has a generating object $X$
whose tensorsquare $X\otimes X$ is  the sum of two simple objects. In this
situation we face a much larger class of categories including those that are
obtained as product-, orbit-, and subgrading-categories from the known ones.
The mentioned constructions rely on the study of gradings and invertible
objects of a BTC.
Many of the resulting categories are inequivalent to any of  the semisimplified
representation categories of Hopfalgebras and those occurring in conformal
field theory. We find a natural condition in terms of Hecke algebra
representations for when this list of categories is complete. We prove it for
the case where one of
the summands of $X\otimes X$ is  invertible, thereby yielding a complete
classification.

\medskip
\noindent
{\bf Acknowledgements :} I thank P.Deligne, J.Fr\"ohlich,
D.Kazhdan, and H.Wenzl for very useful discussions.  This work was in part
supported by NSF grant DMS-9305715.

\setcounter{chapter}{1}
\subsection*{1.Braided Tensorcategories} In all our consideration we mean by a
braided tensorcategory $\cal C$ an {\em abelian category} (see [M]) for which
the morphism sets are finite dimensional vectorspaces over $\bC\,$ . In
addition we have natural
transformations \mbox{$\epsilon\,\in\,Nat(\otimes,P\otimes)$} and
\mbox{$\alpha\,\in\,Nat(\otimes(id\times\otimes),\,\otimes(\otimes\times
id))$}.
They yield the commutativity and associativity isomorphisms
$\;\epsilon(X,Y)\,:\,X~\!\otimes\!~Y\to Y\!\otimes\! X$ and
$\alpha (X,Y,Z):X\!\otimes\!(Y\!\otimes\! Z)\to (X\!\otimes\! Y)\!\otimes\! Z$
which have to obey the pentagonal and two hexagonal equations. For simplicity
we shall omit $\alpha$ in the formulas although it can be a non trivial
morphism. Also we shall
only consider {\em rigid} categories. This means that to any object $X\in {\cal
C}$ we find a conjugate object $X^{\vee}$ and morphisms
\mbox{$ev\,:\,X^{\vee}\!\otimes\! X \to 1 $} and \mbox{$coev\,:\,1\to X
\!\otimes\! X^{\vee}\,$}, with the usual pair of
contraction identities. For details see, e.g., [S] for the symmetric and [K]
for the braided case.

For any $\otimes-$category $\cal C$ we can define the {\em fusionring}
$K^+_o({\cal C})$, which is the ring over $\bZ^+$ generated by the equivalence
classes $[X]$ of objects subject to the relations \mbox{$[X]\,=\,[Y]+[X/Y]$}
whenever $Y$ is included into $X$ and $[X\!\otimes\! Y]=[X][Y]\,$. It is clear
that with this definition every object can be written uniquely as the sum of
the simple objects that appear in its  composition series and the products of
the simple objects determine all other products of the fusionring.

A notion that is very useful for our purposes is that of {\em grading}. For a
BTC the set $\nC$  which consists of natural transformations $\xi (X)\in
End(X)$ with \mbox{$\xi (X\!\otimes\! Y)=$}\mbox{$\xi (X)\otimes \xi (Y)$} is
an abelian group. This fact allows us to decompose every object uniquely into a
direct sum \mbox{$ X\,=\,\bigoplus_{\nu\in Gr({\cal C})}X_{\nu}$}. Here
$X_{\nu}$ is the maximal subobject such that the only eigenvalue of
$\xi(X_{\nu})$ is $\nu (\xi)\,$ for all $\xi\,$.
$Gr({\cal C})$ is the subgroup of all characters on  $\nC$  of
this form. This decomposition has the property that $\bigl(X\!\otimes\!
Y)_{\nu}\,=\,\oplus_{\eta}X_{\nu{\eta}^{-1}}\otimes Y_{\eta}\,$ and that to any
{\em simple} object $X$ we can assign a unique $\nu \in Gr({\cal C})$ with
$X=X_{\nu}$.  Thus $Gr({\cal C})$ makes $\kC$ into a graded algebra. We call
${\cal C}$ {\em locally rational} if every component $\kC_{\nu}$ is finitely
generated , i.e., if there are only finitely many inequivalent, simple objects
of a given grading.

A special type of simple objects are the {\em invertible} ones, which satisfy
$X\!\otimes\! X^{\vee}\,\cong\, 1\,$. They form an abelian group on $\kC$ we
shall call  $\pC$. Let us introduce two natural group homomorphisms:
\begin{equation}\label{homt}
\vartheta \, :\,\pC\longrightarrow \gC\qquad
\end {equation}
 \begin{equation}\label{homm}
\mu\,:\,\pC\longrightarrow \nC\;,
\end {equation}
where $\vartheta$ associates a grading to an irreducible element in $\pC$ and
$\mu$ is defined by
$
1_g\otimes\,\mu(g)\!\bigl(X\bigr)\,=\,\epsilon(X,g)\epsilon(g,X)\,.
$
A {\em balancing} of a tensorcategory is a natural transformation of $X\to
X^{\vee\vee}$ to the identity functor. For a BTC a balancing is equivalently
given by a transformation $\theta \in \rnC$ with
$$
\epsilon(Y,X)\epsilon(X,Y)=\theta(X)\otimes\theta(Y) \theta (X\!\otimes\!
Y)^{-1}\; {\rm and }\;\;\theta(X^{\vee})=\theta(X)^t\;.
$$
If such a balancing exists (there are plenty of examples where it does not)
it is unique up to elements of order two in $\nC\,$.
To a given balancing we can associate a family of traces $tr_X\in End(X)^*$ by
 $$
tr_X(f):\,1\stackrel {coev}{\hbox to 26pt{\rightarrowfill}} X \otimes X^{\vee}
\stackrel {(f\theta(X))\otimes 1}{\hbox to 38pt{\rightarrowfill}}
X \otimes X^{\vee} \stackrel {\epsilon(X , X^{\vee})}{\hbox to
32pt{\rightarrowfill}}X^{\vee} \otimes X \stackrel {ev}{\hbox to
20pt{\rightarrowfill}}1\;.
$$
We call a {\em dimension} a function $d\,:\, \kC\to \bC $ which respects sums
and products and is invariant under conjugation. Since the trace is cyclic,
also for pairs of morphisms between different objects,
and factorizes w.r.t. tensorproducts, we can define a canonical dimension by
$d_{tr}(X)=tr_X(1)\,$.
Dimension functions can also be constructed in a different way by applying
Perron-Frobenius theory to the fusion matrices of
$\kC$, representing the action of the ring on itself by multiplication.
\begin{theorem}\label{PF}
Assume that the fusionring $\kC$ of a BTC $\cal C$ is locally rational, then
\begin{enumerate}
\item there is exactly one \underline{positive} dimension $d_{PF}:\kC\to
{\bR}^+\,$,
\item $d_{PF}\geq 1$ and $d_{PF}(X)=1$ if and only if $X\in \pC \,$.
\item If $X=X_{\eta}$ then $\underline{X}\,:\,\kC_{\nu}\to\kC_{\nu\eta}$,
defined by multiplication has norm $d_{PF}(X)$ independent of $\nu\,$.
\end{enumerate}
\end{theorem}
In the last statement we assumed $\kC$ to be equipped with the inner product
for
which the simple objects are an orthonormal basis.

In order to relate the positivity condition to properties of the categories
themselves we introduce $C^*$ structures  which are known from  applications in
 operator algebras and physics [DR], but are also related to the {\em
polarizations} in [S]. A {\em *-structure } on a BTC is an antilinear,
contravariant, coexact BTC-functor  $*\,:\,{\cal C}\to{\cal C}\,$. For
simplicity
let us assume that $X^*\otimes Y^* \widetilde{\longrightarrow} (X \otimes
Y)^*$ is the identity so that $\alpha$ and $\epsilon$ are unitary. We call
the category of finite dimensional Hilbert spaces ${\cal H}$ and denote by
$\Omega$ the class of all covariant, exact (not necessarily $\otimes-$)
functors $\omega\,:\, {\cal C}\to{\cal H}$ which commute with * . We say that
$\cal C$ is a {\em $C^*$-category}
if for any morphism $f$ there is some $\omega \in \Omega$ with $\omega(f)\neq
0$. In this case we can introduce a norm
$\|f\|=sup_{\omega\in\Omega}\|\omega(f)\|$ which renders the category $\cal C$
semisimple and equips the algebras $End(X)$ with a $C^*$-structure in the usual
sense. For $C^*$-categories we construct a balancing as follows. Define
$\lambda_X\in End(X)$ by
$$
X \stackrel{1\otimes ev^*}{\hbox to 30pt{\rightarrowfill}}X\otimes
X^{\vee}\otimes X  \stackrel{\epsilon(X,X^{\vee})\otimes 1}{\hbox to
38pt{\rightarrowfill}} X^{\vee}\otimes X\otimes X\stackrel{ev\otimes
1}{\hbox to 30pt{\rightarrowfill}} X \;.
$$
 From the positivity of $\,<f>\,=\,ev(1\otimes f)ev^*\,=\,tr_X(\lambda^*_X f)$
and the fact that $End(X)$ is a sum of
type $I$ factors with trace $\,tr_X$ we infer that $\,\lambda_X$ is central.
Since $tr_X$ is generally cyclic it follows that the unitary part $\theta_o(X)=
U(\lambda_X )$ gives rise to a natural transformation.
\begin{theorem}
To any $C^*$-BTC $\,\cal C$ there exists precisely one balancing such that the
associated traces $tr_X$ are positive $\forall X\in ob({\cal C})$. It is given
by $\theta_o\in \rnC\,$.
\end{theorem}
Clearly,  for this choice, the dimension $d_o$ associated to the balancing is
positive. Thus by Theorem \ref{PF} we obtain for locally rational
$C^*$-categories
the remarkable identity
\begin{equation}\label{dim}
d_{PF}\,=\,d_o
\end{equation}
where both quantities are defined in completely independent ways.

\setcounter{chapter}{2}
\subsection*{2.Hecke - and Temperley Lieb Type Categories}
In many examples $\kC$ is generated by a single object $\Pi$ (e.g., a
fundamental representation) meaning every object is the direct sum of
subobjects of tensorpowers of $\Pi\oplus\Pi^{\vee}$. It is easy to see that in
this situation $\gC \, \cong {\bZ} / N$ , generated by the character of $\Pi$,
and the order $N\geq 1 $ is the smallest number such that
$Hom\bigl(\Pi^n,\Pi^{(n+N)}\bigr)\neq 0$ for some $n\,$.

In order to state a tractable classification problem we confine the class of
BTC's further by restricting the dimension of $End(\Pi^{\otimes 2})\,$.
The condition $End(\Pi^{\otimes 2})=\bC$ is  by  rigidity equivalent to $\Pi
\in \pC$ whereas $\,End( \Pi^{\otimes 2})=\bC \oplus \bC$ implies that
$\Pi\otimes\Pi\cong A\oplus B$ for two inequivalent, simple objects $A$ and
$B$.
The first is a special case of  a $\theta$-category which we classify in the
next section. In the second case $\epsilon(\Pi,\Pi)$ has two eigenvalues
$\gamma_A$ and $\gamma_B$ so that the rescaled natural representation of n-th
braidgroup $B_n$
on ${\cal E}_n=End(\Pi^{\otimes n})\,$, defined by
$\rho(g_{i+1})=-\gamma_A1^{\otimes i}\otimes \epsilon(\Pi,\Pi) $ factors into a
representation of the n-th Hecke algebra  $ \rho :\,H_n(q)\to {\cal E}_n$ with
$q:=-\gamma_B\gamma_A^{-1}\,$. (We choose conventions as in [W1].) This
sequence of morphisms is compatible with the inclusions
 ${\cal E}_n \hookrightarrow {\cal E}_{(n+1)}\,:\,f\mapsto f\otimes 1_{\Pi}\,$
and thus extends to $ \rho :H_{\infty}\to {\cal E}_{\infty}$. If $\cal C$ is
also a $C^*$-category we have $|q|=1$ and $\rho$ is a *-representation on every
$H_n(q)$. Henceforth we call BTC's with these properties {\em Hecke type
categories}. For these $\tau|_{{\cal E}_n}\,=\,d(\Pi)^{-n}tr_{\Pi^n}$  defines
a positive, normalized Markov trace on ${\cal E}_{\infty}$ with modulus
$\eta\,=\,\tau(e_A)\,=\,d(A)d(\Pi)^{-2}$.
Combining the above observations with results from [W1] we find the following
restrictions:
\begin{theorem}\label{Hecke}
For a Hecke type category with $\epsilon(\Pi,\Pi)^2$ nonscalar we have
\begin{enumerate}
\item $q\,=\,-\gamma_B\gamma_A^{-1} =e^{\pm \frac {2\pi i} l}$ for some
$l=4,5,\ldots$
\item $\eta=\frac{d(A)}{d(\Pi)^2}=\frac {(1-q^{(-k+1))}}{(1+q)(1-q^{-k})}$
for some $k\,=\,1,\dots,l-1\,$.
\item The morphism $\rho$ factors through the semisimple quotient
 $H_n(q)\to H^{(k,l)}_n$ whose representations are labeled by $(k,l)$-diagrams.
\end{enumerate}
\end{theorem}
Since $H^{(k,l)}$ coincides with the GNS-quotient of the pullback $\rho
^*\tau\,$, the factorized morphism $\bar \rho\,:\,H^{(k,l)}_{\infty}\to {\cal
E}_{\infty}$ is an
inclusion. It also yields a morphism of (non rigid) fusionrings $K_o(\rho
)_n\,:\,K_o^+\bigl(H_n^{(k,l)}\bigr)\to\kC_n$ for positive gradings
$n=0,1,\ldots\,$. Here $K_o^+\bigl(H_{\infty}^{(k,l)}\bigr)$ has a unique,
smallest extension into a rigid fusionring $F^{(k,l)}$ with $\bZ$-grading which
is shown in [GW] to be isomorphic to the truncated subfusionring of
$U_q(Gl(k))$ generated by the usual fundamental representation. If $\cal C$ is
locally rational the norms of $\underline{[1]}|_{H_n^{(k,l)}}$ and
and $\underline{\Pi}|_{\kC_n}$ and hence of $\| K_o(\rho)_n\|$ are independent
of $n$ for large $n\,$. In this situation we find that $K_o(\rho)([1^k])$ has
norm one,
i.e., it is invertible, and thus can be used  to extend $K_o(\rho)$ to
a morphism of rigid fusionrings $\Psi\,:\,F^{(k,l)}\to \kC\,$, defined also for
negative gradings.

The embedding of the Hecke algebras gives us not only information on the
fusionring but allows us  to compute the balancing phases.
In $H(q)$ the scalar $\alpha_{\lambda}$ by which the
central braid group element $\Delta_N^2\,=\,\bigl(g_1\ldots g_{(N-1)}\bigr)^N$
acts in the irreducible representation associated to the diagram
 $\lambda$
has been computed in [W2] as a framing anomaly of link invariants.
It is possible to factorize the product of $\epsilon$'s in ${\cal E}_N$
associated to $\Delta_N^2$ into the expression $\theta(\Pi)^{\otimes N}
\theta(
\Pi^{\otimes N})^{-1}$. This observation enters the second part of the
following theorem.
\begin{theorem}\label{phi}
If $\cal C$ is a locally rational Hecke type category, then
\begin{enumerate}
\item there is a unique morphism of rigid fusionrings
$$
\Psi\,:\,F^{(k,l)}\longrightarrow \kC \quad {\rm with} \; \Psi([1])=\Pi
$$
\item if $X\in ob({\cal C})$ is a subobject of $\Psi(\lambda)$ for some
diagramm $\lambda$ then
$$
\theta(X)\,=\,\theta_{\lambda}1_X \quad {\rm where}
\quad\theta_{\lambda}=\theta(\Pi)^{|\lambda |}(-\gamma_A)^{|\lambda |-|\lambda
|^2}\alpha^{-1}_{\lambda}
$$
\end{enumerate}
\end{theorem}
 By definition the image of $\Psi$ generates additively  $ob({\cal C})$ so that
every object is  a sum of those considered in $b)$. Hence the balancing of a
Hecke type category is completely determined by $\,\theta(\Pi)$, $\,\gamma_A$
and $\gamma_B\,$.

 In order to explain the constraints on $\Psi$ resulting from
Theorem (\ref{phi}) we define the graph of a map of positive lattices
$\Lambda\,:\,L_1 \to L_2$ as the bicolored graph whose vertices are the
generators of $L_1$ and $L_2$ with respective coloration. The number of edges
between them are given by the matrix elements of $\Lambda\,$. Denoting by
$\Psi_n$, $\underline{[1]}_n$ and $\underline{\Pi}_n$ the respective
restrictions to the $n$-th graded components
the relation $\Psi_{(n+1)}\underline{[1]}_n\,=\,\underline{\Pi}_n\Psi_n$ means
that pairs of neighboring simple objects in the graph of $\underline{[1]}_n$
are mapped by $\Psi$ to sums of pairs of neighboring objects in the graph of
$\underline{\Pi}_n\,$. By Theorem \ref{PF} $\Psi$ is dimension preserving,
i.e., $d_{PF}(\Psi(X))=d_{PF}(X)\,$.
 From part $b)$ of Theorem \ref{phi} we see that $\theta$ has to have the same
value on every
simple object of a connected component of the graph of $\Psi_n\,$. Knowing the
specific values for one coloration namely the $\theta_{\lambda}$ on $F^{(k,l)}$
this imposes together with the neighborhood condition strong constraints on the
structure of $\Psi_n$. In many cases the only remaining possibility is that the
components of $\Psi_n$ are pairs of different colorations so that   $\Psi_n$ is
an isomorphism for every $n\,$.
In this case we say that $\Psi$ is a {\em local isomorphism}. A special
subclass of such categories are {\em Temperley Lieb type categories} which are
defined in the next theorem. Its proof is in part a direct consequence of
Theorem \ref{PF} and identity (\ref{dim}).
\begin{theorem}\label{TLcats}
If $\cal C$ is a Hecke type category with $\epsilon(\Pi,\Pi)^2$  nonscalar,
then the following four conditions are equivalent
\begin{tabbing}
1.) $k=2$\hphantom{xxxxxxxxxxxx} \= 2.) $e_A$ and $1\!\otimes\! e_A$ generate
${\cal A}_{\beta}(3)$ with $\beta < 4$\\
3.) $A\,\in\, Pic({\cal C})$ \> 4.) $d(X) < 2\,$ and $\,d(A)\leq d(B)
\,$.
\end{tabbing}
\end{theorem}
Here ${\cal A}_{\beta}(n)$ is the Temperley Lieb quotient of the Hecke algebra
with modulus $\beta=q+q^{-1}+2=d(\Pi )^2\,$.
 The elements of $F^{(2,l)}$ are
pairs $[\lambda_1,\lambda_2]$ with $\lambda_i\in\bZ$ and
$0\leq\lambda_1-\lambda_2\leq l-2\,$. The graph associated to $[1]_n$ is
$A_{l-1}\,$, where the gradation is $n= \lambda_1+\lambda_2$ and two simple
objects are adjacent if they coincide in one component. Specializing Theorem
{}~\ref{phi} to $k=2$ we can write the balancing as
$\theta_{\lambda}=c_nt^{d^2}$
where $t^4=q$ is the primitive $l$-th root of unity and
$d=\lambda_1-\lambda_2+1\,$.

The norm of $\underline{\Pi}_n$ has to be the same as the norm of
$\underline{[1]}_n$ so that
by an old result of Kronecker, see [GHJ], the only possibilities for the graph
of $\underline{\Pi}_n$ are $A_{l-1}\,$,
$D_{l/2+1}\,$ or $E_{6,7,8}\,(l=12,18,30)\,$. One readily checks that the
neighborhood and component condition discussed above exclude the $D$ and $E$
cases. In summary, we have the following result for $k=2\,$:
\begin{theorem}\label{locis}
Suppose $\cal C$ is a Temperley Lieb type category  with \mbox{$\beta\neq 4$}.
Then there exists a local isomorphism
of rigid, graded fusionrings \quad $\Psi\,:\,F^{(2,l)}\to\kC\,$ with
$\Psi([1])\,=\,\Pi\,$.
\end{theorem}
\setcounter{chapter}{3}
\subsection*{3. Two Important Examples}
A.) A class of braided tensorcategories that can be completely classified are
semisimple BTC's for which all simple objects are invertible. We call them {\em
$\theta$-categories}. For a $\theta$-category $\cal C$ we have in particular
$\kC\cong\bZ^+[\pC]\,$ and the map  $\vartheta$ from (\ref{homt}) yields an
isomorphism $\,\pC\cong\gC\,$. To a class of pairs $(\epsilon,\alpha)$ of
natural isomorphisms (considered as functions $\alpha\in C\bigl(\pC^3\bigr)$
and $\epsilon\in C\bigl(\pC^2\bigr)$ by specialization) that give rise to
equivalent BTC's we can assign a unique class in $H^4\bigl(G,2;\bC^*\bigr)\,$,
the cohomology group of the Eilenberg MacLane space $K_2(G)\,$.
This correspondence results from the fact that the pentagonal and hexagonal
equations translate to cocycle conditions and the transformations $g\otimes
h\widetilde{\rightarrow}g'\otimes h'$ of $\otimes$-isomorphisms give rise to
coboundaries, see [FK]. The function
$\,\theta:\pC\to\bC^*\,;\,g\mapsto\epsilon(g,g)\,$ is easily shown to be
quadratic, only dependent on the cohomology class of $\epsilon$ and a possible
balancing of $\cal C\,$. Combining these observations with results in [EM]
we find the following classification:
\begin{theorem}\label{theta}
To any quadratic form $\theta$ on a finitely generated abelian group $G$ there
exists one and up to isomorphism only one $\theta$-category ${\cal
P}(\theta,G)$ such that $Pic\bigl({\cal P}(\theta,G)\bigr)\cong G$ and $\theta
(g)=\epsilon(g,g)\,$.
\end{theorem}
B.) It is well known that the category $Rep\Bigl(U_t\bigl(Sl(k)\bigr)\Bigr)\,$
of quantum group representations, with $q=t^{-2k}$ a primitive $l$-th root of
unity, is not semisimple. Nevertheless, it is possible to define a semisimple
subquotient category. The morphisms are the quotients
of $Hom_{\cal C}(X,Y)$ by the nullspaces $Hom(X,Y)^o$ of the trace pairing
$$
Hom(Y,X)\otimes Hom(X,Y)\,{\hbox to 30pt{\rightarrowfill}}\, End(X)
\stackrel{tr_X}{\hbox to 30pt{\rightarrowfill}} \bC^*
$$
 In this category we also discard objects with $End(X)=End(X)^o$ which for
indecomposable $X$ is equivalent to $d(X)=0\,$.(For details of this
construction see [K] and also [A] and [GK].) The full subcategory generated by
the image of $\Pi=[1]$ is a
semisimple Hecke type category $\,R(t,k)\,$ without an apriori *-structure.
Let us call this an {\em indefinite Hecke type category}. As for $Sl(k)$ we
label the simple objects by Young diagrams with the restrictions $0 \leq
\lambda_1-\lambda_k\leq l-k$ so that $A=[1,1]\,$, $B=[2]\,$,
$\gamma_A=-t^{1+k}$ and $\gamma_B=t^{1-k}\,$. The group
$Pic\bigl(R(t,k)\bigr)\cong\bZ/k$ is generated by the $\alpha=[l-k]\,$.
The grading group $Gr\bigl(R(t,k)\bigr)$ is also cylic of order $k$ and
associates to a diagram $\lambda$ the number of boxes $|\lambda|\, mod\,k\,$.
Hence $\vartheta:\bZ/k\to\bZ/k$ from (\ref{homt}) is just multiplication with
$l\,$. A possible balancing of $\,R(t,k)\,$ is given by
$\theta_{\lambda}=t^{c(\lambda)}\,$ where $$
\,c(\lambda)=\sum_{i<j}(\lambda_i-\lambda_j)^2+k(\lambda_i-\lambda_j)\,.
$$
 The structure of the full subcategory over $Pic({\cal C})$ is determined in
the sense of sense of Theorem~\ref{theta} by
$\,\epsilon(\alpha,\alpha)=(-1)^{(l-k)}t^{(l-k)l}\,$. The map $\mu$ defined in
(\ref{homm}) is given by $\,\mu(\alpha)\bigl([1]\bigr)=t^{2l}\,$.
A deformation argument used in [FK] (which should be extendable to general $k$)
shows that the necessary constraint in Theorem \ref{Hecke} for the existence of
*-structures is also sufficient:
\begin{theorem}\label{def}
$\,R(t,2)\,$ is isomorphic to a $C^*$-category if and only if
$t^4=e^{\pm\frac{2\pi i}l}\;$.
\end{theorem}
There is a remarkable uniqueness result on the categories with the same fusion
ring as $R(t,k)\,$ due to [KW] (for a proof
for $k=2$ using structure constants see [FK]).
\begin{theorem}\label{slk}
Suppose for an indefinite Hecke type category $\cal C$ there is an isomorphism
of fusionrings $\psi\,:\,\kC\widetilde{\to}K^+_o\bigl(R(t,k)\bigr)$ mapping
generators to each other. If in addition the invariants $\gamma_A\,$ and
$\gamma_B\,$ of $\cal C$ coincide with those of $R(t,k)$ then $\psi$ extends to
an isomorphism of categories ${\cal C}\cong R(t,k)\,$.
\end{theorem}
\setcounter{chapter}{4}
\subsection*{4. Product and Orbit Categories}
There are a number of natural operations between categories that allow us to
produce new categories, e.g., from the examples in the previous section. A
special class of $\otimes-$ subcategories of a given BTC $\cal C$ is obtained
by picking
a subgroup $H\subset Gr({\cal C})$ and defining $_H\!{\cal
C}\hookrightarrow{\cal C}$ to be the largest full subcategory for which all
objects have grading in $H\,$. Of particular interest is the subcategory
$_0{\cal C}$ which consists of objects with trivial grading. It is additively
generated by the subobjects of all $j\otimes j^{\vee}$ with $j$ simple. Also we
denote by $\,{\cal C}_1\cap{\cal C}_2\,$ the largest full subcategory which is
contained in two full $\otimes-$subcategories $\,{\cal C}_i\hookrightarrow
{\cal C}\,$.

Dual to the notion of direct products of Hopfalgebras we have the notion of a
product of categories ${\cal C}_i$ which is a biexact functor
\mbox{$\odot\,:\,{\cal C}_1\times {\cal C}_2\to {\cal C}_1\odot{\cal C}_2\,$}
onto the smallest additive completion of the ordinary product. The precise
definition is given in [D]. Clearly, this functor induces an isomorphism
$Gr\bigl({\cal C}_1\bigr)\oplus Gr\bigl({\cal C}_2\bigr)\,\cong\,Gr\bigl({\cal
C}_1\odot  {\cal C}_2\bigr)\,$.

The notion of quotients of BTC's  related to branching of representations to
sub-Hopfalgebras needs more explanation: To this end assume that $P$ is a
full $\otimes-$subcategory with a $\otimes-$fibre functor $\nu\,:\,P\to
Vect(\bC )$ (or $\cal H$) of strict, symmetric categories.  To any object $X\in
ob({\cal C})$ we have - up to isomorphism - a unique maximal subobject
$X_P\hookrightarrow X$ with $X_P\in ob(P)\,$. We define a category ${\cal C}/P$
with $ob\bigl({\cal C}/P\bigr)=ob\bigl({\cal C}\bigr)\,$ and morphisms
$\widetilde{Hom}(X,Y)=\nu\bigl((Y\otimes X^{\vee})_P\bigr)\,$.
(see [D],[DM] for Tannakian categories.)
The canonical morphism in~$P$,  $(Z\otimes Y^{\vee})_P\otimes (Y\otimes
X^{\vee})_P\,\to\,(Z\otimes X^{\vee})$, obtained from $ev\,$,
 determines the composition of morphisms in ${\cal C}/P\,$. Using the natural
braid isomorphisms we find two canonical isomorphisms in $P$
\begin{equation}\label{mort}
\oslash^{\pm}\,:\,(Y_1\otimes X_1^{\vee})_P\otimes (Y_2\otimes
X_2^{\vee})_P\,\to\, \bigl((Y_1\otimes Y_2)\otimes(X_1\otimes
X_2)^{\vee}\bigr)_P
\end{equation}
both of which define tensorproducts of morphisms in  ${\cal C}/P\,$.

Viewing the invariances as subobjects $\,Z_1\hookrightarrow Z_P\,$ the map

\noindent
$
 Hom(X,Y) \to Hom\bigl(1,(Y\!\otimes\!X^{\vee})_1\bigr)\,\stackrel {\nu}{\hbox
to 15pt{\rightarrowfill}}\,$
$
Hom_{\bC}\bigl(1,\nu\bigl((Y\!\otimes \!X^{\vee})_1\bigr)\bigr)\,\widetilde
{\hbox to 15pt{\rightarrowfill}}\;$
 \mbox{$ \nu\bigl((Y\!\otimes\! X^{\vee})_1\bigr) \hookrightarrow\;$}
\mbox{$ \nu\bigl((Y\!\otimes \!X^{\vee} )_P\bigr)\,$}
gives then rise to a $\otimes-$functor $\,p:{\cal C}\to{\cal C}/P\,$ such that
the following diagram commutes:
\begin{equation}\label{comm}
\begin{array}{ccl}
\raisebox{.4ex}{${\cal C}$}\, &\raisebox{.4ex}{$\stackrel{p}{\hbox to
40pt{\thrafill}}$}&\;\raisebox{.4ex}{${\cal C}/P$}\\
\hup & & \quad \hup  \otimes 1_{\cal C}\\
P\;&\stackrel{\nu}{\hbox to 40pt{\thrafill}}&\,Vect(\bC)
\end{array}
\end{equation}
Clearly, the images of the natural isomorphisms $\bar{\epsilon}=p(\epsilon)$
and $\bar{\alpha}=p(\alpha)$ satisfy the pentagonal and hexagonal equations and
are natural with respect to morphisms in the image of $p\,$. But since the
functor $p$ is by definition not full for $P\neq Vect(\bC)$ there is a priori
no reason for $\bar{\epsilon}$ and
$\bar{\alpha}$ to be natural in ${\cal C}/P\,$. It turns out that naturality is
equivalent to demanding that $P$ {\em decouples}, i.e.,
$$
\epsilon(Q,X)\epsilon(X,Q)\,=\,1 \quad {\rm for\;\, all } \quad Q\in
ob(P),\,X\in ob({\cal C})\;.
$$
In this case the two morphisms $\oslash^{\pm}$ from (\ref{mort}) coincide.
Suppose $j\in ob({\cal C})$ is simple and $j\otimes j^{\vee}$ contains
nontrivial subobjects from $P\,$. Then $\widetilde{End}(j)\neq \bC\,$, and
since kernels and cokernels have to stem from $\cal C$,   ${\cal C}/P\,$ fails
to be abelian. Also na\"{\i}ve abelian completions usually spoil naturality of
$\bar{\epsilon}\,$. In order to avoid this situation we have to impose the
condition ${\cal C}_o\cap P\,=\,Vect(\bC )\!\otimes\!1\,$. It is easily seen
that the only subcategories with this property are $\theta$-categories over
subgroups $R\subset Pic({\cal C})$  on which the grading $\vartheta |_R$ from
(\ref{homt}) is injective. In this case the fusionring morphism associated to
$p$ is locally isomorphic and the inequivalent objects of ${\cal C}/P$ are
identical with orbits of $R$. Hence we call ${\cal C}/R$ an {\em orbit
category}. (In [FK] the term {\em induced category} was used.)

 Conversely, any local isomorphism $\psi$ is of the form that it sends simple
objects to their orbits under the action of $\psi^{-1}(1)\subset \pC \,$.
Moreover, we can pullback every category along such $\psi$ by setting
$$
Hom_{\cal C}(X,Y):=\bigoplus_{\nu\in Gr({\cal C})} Hom_{{\cal
C}/P}(\psi(X_{\nu}),\psi(Y_{\nu}))\,.
$$
 Note that the decoupling condition for $R$ is that $R$ lies in the kernel of
the map $\mu$ from (\ref{homm}). We conclude with a survey of properties of
orbit categories. For more details see [FK].
\begin{theorem}\label{orbit}
\begin{enumerate}
\item If $R\subset\pC$ is a subgroup on which $\mu$ is trivial, $\vartheta$ is
injective and the associsted $\theta$-subcategory $P$ is trivial then there
exist  a unique, abelian BTC ${\cal C}/P\,$,  and functors $\nu$ and $p$
such that (\ref{comm}) commutes.
\item For any local isomorphism $\psi :F\to K^+_o(\bar{\cal C})\,$ of rigid
fusionrings there is   a unique BTC $\cal C$ with $\kC\cong F$, a functor
$p:{\cal C}\to \bar{\cal C}$ and a fibre functor on the subcategory associated
to $\psi^{-1}(1)$ extending $\psi$ such that (\ref{comm}) commutes.
\end{enumerate}
\end{theorem}
\setcounter{chapter}{5}
\subsection*{5. A New Family of Hecke  Categories and a\\
Classification of
Temperley Lieb Categories}
Combining the constructions and examples given in the previous sections we can
define a class of indefinite Hecke type categories with fusionring $F^{(k,l)}$
by
$$
D'(\theta,t,k)\,:=\, _{\Delta}\Bigl({\cal P}(\theta,\bZ)\odot D(t,k)\Bigr)
$$
where $\Delta\subset \bZ\oplus\bZ/k = Gr({\cal P}\odot  D)$ is the diagonal
subgroup. The basic invariants with respect to the canonical generator
$\Pi'=(1)\!\odot\!\Pi$ are $\gamma_A=-\theta(1) t^{1+k}$ and
$\gamma_B=\theta(1) t^{1-k}\,$, where $(1)$ is the generator of $\cal P\,$.
 In fact Theorem \ref{slk} and Theorem \ref{orbit} show that $D'$ is the only
category with this fusionring and
these invariants. We have an isomorphism
$$
\varphi\,:\,\bZ\oplus\bZ/(k,l)\widetilde{\hbox to 20pt{\rightarrowfill}}
Pic(D')\,;\,(i,j)\mapsto
((k,l)i)\odot\alpha^{jk'+il''}\;,
$$
 where $k'=k/(k,l)$ and $ll''=(k,l)\,mod\,k\,$. The grading $\vartheta$ is the
projection onto the first factor $i(k,l)\,$.
The $\theta$-category $P_{ij}$ associated to the infinite cyclic subgroup
generated by an object $\varphi(i,j)=(n)\odot\alpha^m$ with $n\neq 0$ is
trivial and decouples iff
$\epsilon(\alpha,\alpha)^{m^2}={\theta(1)}^{-n^2}\,$ and
$t^{2lm}={\theta(1)}^{2n}\,$. For these values we denote by
\begin{equation}\label{twist}
D''(\theta,t,k,i,j):=D'(\theta ,t,k)/P_{ij}\,
\end{equation}
the orbit category as defined in Theorem \ref{orbit}.  In the list of the
categories of the form (\ref{twist}) we recover the ones obtained from
$\widehat {s\ell}(k)_{l-k}$ and $\widehat {s\ell}(l-k)_{k}$ and products of
these with level one theories. Using that the group extension
$$
0\,{\longrightarrow} Pic( _0{\cal C})\,{\longrightarrow} Pic({\cal C})
\,{\longrightarrow}Gr({\cal C})
$$
is an invariant of $\cal C$ we can easily check that the orbit construction
yields categories inequivalent to any subcategories of the known
representation categories of Hopfalgebras. The easiest such case is found for
$l=6,\,k=2$, if we divide by the $\theta$-subcategory generated by
$\varphi(1,1)=(2)\odot [4]\,$. The set of simple objects
$\,\{\,[\cdot],\dots,\, [4]\}\,$ is the same as for the $U_t(sl_2)$ category,
but we have modifies products $[1][1]=[3][3]=[2]+[4]\,$ and
$[1][3]=[\cdot]+[2]\,$.
In general the requirement {\it 2.)} of local isomorphie from Theorem
\ref{orbit} is difficult to verify. However for $\, k=2\,$ we can use Theorem
\ref{locis}
and the uniqueness of the $D'$-categories to prove the following
classification.
\begin{theorem}\label{class}
Every Temperley Lieb type category with $\epsilon(\Pi,\Pi)^2$ nonscalar is of
the form $D''(\theta,t,2,i,j)$ for $t^4=e^{\pm\frac {2\pi i}l}$ and admissible
$\theta$, $i$ and $j\,$.
\end{theorem}
In the case where $\epsilon^2$ is scalar we can consider products with suitable
$\theta$-categories and reduce the problem to the case where
$\epsilon(\Pi,\Pi)^2=1\,$.
Since $\Pi$ is a generator this implies that the category is symmetric and we
can apply the result of [DR] to find a classification in terms of
$U(2)$-subgroups.
\subsection*{References}
\noindent
{[A]}Anderson,H.: Tensor products of quantized tilting modules. Commun. Math.
Phys. {\bf 149}, 149-159(1991).

\noindent
{[D]}Deligne,P.:Cat\'egories Tannakiennes. Grothendieck Festschrift, {\bf 2},
111-195.  Birkh\"auser, 1991. Deligne,P.,Milne,J.: Tannakian categories. Lect.
Notes in Math. {\bf 900}. Springer Verlag, 1967.

\noindent
{[DR]}Doplicher,S.,Roberts,J.: A New Duality for Compact Groups. Invent. Math.
{\bf 98}, 157-218 (1989).

\noindent
{[EM]}Eilenberg,S.,MacLane,S.: On the Groups $H(\pi,n)$. Ann. of Math. {\bf
58},55-106(1953); Ann. of Math. {\bf 60},49-139(1954).

\noindent
{[FK]} Fr\"ohlich,J.,Kerler,T.: Quantum Groups, Quantum Categories
 and Quantum Field Theory. Lect. Notes in Math. {\bf 1542}.
 Springer Verlag, 1993.
Kerler,T.: Quantum Groups, Quantum Categories and Quantum
\quad Field Theory.Dissertation ETH Nr.9828, 1992.

\noindent
{[GK]}Gelfand,S.,Kazhdan,D.: Examples of Tensorcategories. Inv. Math. {\bf
109}(1992)

\noindent
{[GHJ]}Goodman,F.M.,de la Harpe,P.,Jones,V.F.R.:Coxeter Graphs and Towers of
Algebras. Springer Verlag (1989).

\noindent
{[GW]}Goodman,F.,Wenzl,H.: Littlewood-Richardson Coefficients for Hecke
Algebras at Roots of Unity. Adv. Math. {\bf 82} 244-265(1990).

\noindent
{[KW]}Kazhdan,D.,Wenzl,H.: Reconstructing Monoidal Categories. Adv. Soviet
Math. {\bf 16} 111-136(1993).

\noindent
{[K]}Kerler,T.: Non-Tannakian Categories in Quantum Field Theory. Proceedings
 of the Carg\`ese Summer Institute on "New Symmetries in Quantum
 Field Theory", Carg\`ese, July15-27. Plenum Press, 1991.

\noindent
{[M]}MacLane,S.: Categories for the Working Mathematician. Springer Verlag
1971.

\noindent
{[S]} Saavedra Rivano,N.: Cat\'egories Tannakiennes. Lect. Notes in
Math.,{\bf265}. Springer Verlag, 1972.

\noindent
{[W1]}Wenzl,H.: Hecke Algebras of Type $A_n$ and Subfactors. Invent. Math. {\bf
92} 349-383(1988).

\noindent
{[W2]}Wenzl,H.: Braids and Invariants of 3-Manifolds. To appear in Invent.
Math. (1993).

\newpage

\end{document}